\PassOptionsToPackage{unicode}{hyperref}
\PassOptionsToPackage{hyphens}{url}
\PassOptionsToPackage{dvipsnames,svgnames,x11names}{xcolor}
\documentclass[
]{article}
\usepackage{amsmath,amssymb}
\usepackage{lmodern}
\usepackage{iftex}
\ifPDFTeX
  \usepackage[T1]{fontenc}
  \usepackage[utf8]{inputenc}
  \usepackage{textcomp} 
\else 
  \usepackage{unicode-math}
  \defaultfontfeatures{Scale=MatchLowercase}
  \defaultfontfeatures[\rmfamily]{Ligatures=TeX,Scale=1}
\fi
\IfFileExists{upquote.sty}{\usepackage{upquote}}{}
\IfFileExists{microtype.sty}{
  \usepackage[]{microtype}
  \UseMicrotypeSet[protrusion]{basicmath} 
}{}
\makeatletter
\@ifundefined{KOMAClassName}{
  \IfFileExists{parskip.sty}{%
    \usepackage{parskip}
  }{
    \setlength{\parindent}{0pt}
    \setlength{\parskip}{6pt plus 2pt minus 1pt}}
}{
  \KOMAoptions{parskip=half}}
\makeatother
\usepackage{xcolor}
\providecommand{\tightlist}{%
  \setlength{\itemsep}{0pt}\setlength{\parskip}{0pt}}
\NewDocumentCommand\citeproctext{}{}
\NewDocumentCommand\citeproc{mm}{%
  \begingroup\def\citeproctext{#2}\cite{#1}\endgroup}
\makeatletter
 \let\@cite@ofmt\@firstofone
 \def\@biblabel#1{}
 \def\@cite#1#2{{#1\if@tempswa , #2\fi}}
\makeatother
\newlength{\cslhangindent}
\newlength{\csllabelwidth}
\newenvironment{CSLReferences}[2] 
 {\begin{list}{}{%
  \setlength{\itemindent}{0pt}
  \setlength{\leftmargin}{0pt}
  \setlength{\parsep}{0pt}
  \ifodd #1
   \setlength{\leftmargin}{\cslhangindent}
   \setlength{\itemindent}{-1\cslhangindent}
  \fi
  \setlength{\itemsep}{#2\baselineskip}}}
 {\end{list}}
\usepackage{calc}

\ifLuaTeX
\usepackage[bidi=basic]{babel}
\else
\usepackage[bidi=default]{babel}
\fi
\babelprovide[main,import]{american}

\def\languageshorthands#1{}
\ifLuaTeX
  \usepackage{selnolig}  
\fi
\IfFileExists{bookmark.sty}{\usepackage{bookmark}}{\usepackage{hyperref}}
\IfFileExists{xurl.sty}{\usepackage{xurl}}{} 
\hypersetup{
  pdftitle={SpecpolFlow: a new software package for spectropolarimetry
using Python},
  pdfauthor={Colin P. Folsom, Christiana Erba, Veronique Petit, Shaquann
Seadrow, Patrick Stanley, Tali Natan, Bonnie Zaire, Mary E. Oksala,
Federico Villadiego Forero, Robin Moore, Marisol Catalan Olais},
  pdflang={en-US},
  colorlinks=true,
  linkcolor={Maroon},
  filecolor={Maroon},
  citecolor={Blue},
  urlcolor={Blue},
  pdfcreator={LaTeX via pandoc}}

\title{SpecpolFlow: a new software package for spectropolarimetry using
Python}

\definecolor{c53baa1}{RGB}{83,186,161}
\definecolor{c202826}{RGB}{32,40,38}

\usepackage[affil-it]{authblk}
\usepackage{orcidlink}
\author[1%
  \ensuremath\mathparagraph]{Colin P. Folsom%
    \,\orcidlink{0000-0002-9023-7890}\,%
    }
\author[2%
  ]{Christiana Erba%
    \,\orcidlink{0000-0003-1299-8878}\,%
      \thanks{Co-first author}%
  }
\author[3%
  ]{Veronique Petit%
    \,\orcidlink{0000-0002-5633-7548}\,%
    }
\author[3%
  ]{Shaquann Seadrow%
    \,\orcidlink{0009-0002-0308-2497}\,%
    }
\author[3%
  ]{Patrick Stanley%
    \,\orcidlink{0000-0002-0378-0140}\,%
    }
\author[3%
  ]{Tali Natan%
    \,\orcidlink{0000-0002-7703-6701}\,%
    }
\author[4%
  ]{Bonnie Zaire%
    \,\orcidlink{0000-0002-9328-9530}\,%
    }
\author[5,6%
  ]{Mary E. Oksala%
    \,\orcidlink{0000-0003-2580-1464}\,%
    }
\author[3%
  ]{Federico Villadiego Forero%
    }
\author[3%
  ]{Robin Moore%
    }
\author[3%
  ]{Marisol Catalan Olais%
    \,\orcidlink{0009-0006-2442-6235}\,%
    }

\affil[1]{Tartu Observatory, University of Tartu, Observatooriumi 1,
61602, Toravere, Estonia%
  }
\affil[2]{Space Telescope Science Institute, 3700 San Martin Drive,
Baltimore, MD 21218, USA%
  }
\affil[3]{Department of Physics and Astronomy, Bartol Research
Institute, University of Delaware, 19716, Newark, DE, USA%
  }
\affil[4]{Universidade Federal de Minas Gerais, Belo Horizonte, MG,
31270-901, Brazil%
  }
\affil[5]{Department of Physics, California Lutheran University, 60 West
Olsen Road, 91360, Thousand Oaks, CA, USA%
  }
\affil[6]{LESIA, Observatoire de Paris, PSL University, CNRS, Sorbonne
Université, Université Paris Cité, 5 place Jules Janssen, 92195 Meudon,
France%
  }
\affil[$\mathparagraph$]{Corresponding author: %
}
\date{25 October 2024}

\begin{document}
\maketitle

\section{Summary}\label{summary}

Spectropolarimetry, the observation of polarization and intensity as a
function of wavelength, is a powerful tool in stellar astrophysics. It
is particularly useful for characterizing stars and circumstellar
material, and for tracing the influence of magnetic fields on a host
star and its environment. Maintaining modern, flexible, and accessible
computational tools that enable spectropolarimetric studies is thus
essential. The \texttt{SpecpolFlow} package is a new, completely
Pythonic workflow for analyzing stellar spectropolarimetric
observations. Its suite of tools provides a user-friendly interface for
working with data from an assortment of instruments and telescopes.
\texttt{SpecpolFlow} contains tools for spectral normalization and
visualization, the extraction of Least-Squares Deconvolution (LSD)
profiles, the generation and optimization of line masks for LSD
analyses, and the calculation of longitudinal magnetic field
measurements from the LSD profiles. It also provides Python classes for
the manipulation of spectropolarimetric products. The
\texttt{SpecpolFlow} website includes an array of tutorials that guide
users through common analytic analysis using the software.
\texttt{SpecpolFlow} is distributed as a free, open-source package, with
fully documented tools (via an API and command line interface) which are
actively maintained by a team of contributors.

\section{Statement of need}\label{statement-of-need}

Spectropolarimetry is an essential observational technique in stellar
astrophysics, which is used to study the surface properties of stars and
their environments. Symmetry-breaking phenomena like stellar magnetic
fields leave an imprint on polarized spectra. Magnetic fields are
present in most classes of stars throughout their evolution
(\citeproc{ref-DonatiLandstreet2009}{Donati \& Landstreet, 2009};
\citeproc{ref-Mestel2012}{Mestel, 2012}). Cool stars (with convective
envelopes) display a wide range of observed field strengths, driven by
variations in their internal dynamos
(\citeproc{ref-Reiners2012}{Reiners, 2012}). Nearly 10\% of hot stars
(with radiative envelopes) also harbour strong magnetic fields
(\citeproc{ref-Grunhut2017}{Grunhut et al., 2017};
\citeproc{ref-Sikora2019}{Sikora et al., 2019}). Furthermore, magnetic
fields are found in evolved giants and compact stellar remnants such as
white dwarfs and pulsars (\citeproc{ref-Ferrario2015}{Ferrario et al.,
2015}). Spectropolarimetry is a valuable tool for characterizing the
strength, orientation, and topology of these fields. Therefore the
maintenance and dissemination of computational tools enabling this
technique are a key element of research.

Spectropolarimetric studies of stellar magnetic fields typically
leverage the splitting of spectral lines due to the Zeeman effect (e.g.,
\citeproc{ref-Landstreet2015}{Landstreet, 2015}). The Zeeman split
components of a line are polarized and shifted in wavelength
proportional to field strength. These shifts in wavelength are typically
undetectable due to other line broadening processes (except for very
strong fields); however the changes in polarization remain detectable.
In practical observations this polarization signal is often below the
noise level for an individual line, thus it is important to combine
information from many spectral lines. The Least-Squares Deconvolution
(LSD, \citeproc{ref-Donati1997}{Donati et al., 1997};
\citeproc{ref-Kochukhov2010}{Kochukhov et al., 2010}) approach is the
most widely used method for detecting such polarization signatures. LSD
is a multi-line technique that produces a pseudo-average line profile at
increased signal-to-noise. LSD models the spectrum as the convolution of
a set of delta functions (the `line mask') with a common line shape (the
`LSD profile'). This model is fit to an observation, using a weighted
linear least-squares approach, to derive the LSD profile. An estimate of
the surface averaged line-of-sight component of the field (the
`longitudinal field', \(\langle B_z \rangle\)) can then be computed from
the circular polarization and intensity LSD profiles. Modelling the
variation of \(\langle B_z \rangle\) as the star rotates enables further
characterization, such as determining the stellar rotational period and
simple models of the magnetic topology.

Several successful programs supporting spectropolarimetric analyses
exist in the literature, although they are often not open source and
poorly documented. There is currently no publicly available software
that provides the full toolset needed to analyze reduced
spectropolarimetric observations and produce magnetic field
measurements. The original LSD code by Donati et al.
(\citeproc{ref-Donati1997}{1997}) is efficient and effective, written in
C, but it is both proprietary and undocumented. A more recent program,
\texttt{iLSD,} is an Interactive Data Language (IDL) interface around a
Fortran core implementing LSD (\citeproc{ref-Kochukhov2010}{Kochukhov et
al., 2010}). This code includes additional features such as the
reconstruction of multiple line profiles from the same spectrum, or an
optional Tikhonov regularization of the LSD profiles. However
\texttt{iLSD} is also proprietary and has very limited documentation.
Some additional support codes (e.g., to read and write LSD profiles, to
create line masks, or to combine observations) have been written in a
variety of languages and passed down from person to person. While these
codes are scientifically relevant, they often lack documentation,
version control, active maintenance, and generally they are neither
publicly available nor open source. Factors like these negatively impact
accessibility, serving as a significant barrier to both learning and
research reproducibility.

\section{Overview of SpecpolFlow}\label{overview-of-specpolflow}

The \texttt{SpecpolFlow} package is a modernized, unified revitalization
of the tools that have preceded it. The software is open source, well
documented, and the code itself is extensively commented and designed to
be readable. \texttt{SpecpolFlow} produces consistent results with
previous proprietary codes implementing similar algorithms. It produces
consistent LSD profiles with the code of Donati et al.
(\citeproc{ref-Donati1997}{1997}) and \texttt{iLSD}, and it produces
consistent \(\langle B_z \rangle\) values with the code of Wade et al.
(\citeproc{ref-Wade2000}{2000}).

\texttt{SpecpolFlow} provides a toolkit with an ensemble of Python
functions that:

\begin{itemize}
\tightlist
\item
  Convert observed spectra into a common file format
\item
  Continuum normalize spectra, with an interactive graphical interface
\item
  Generate line masks for LSD from lists of atomic transitions
\item
  Clean line masks interactively to remove problem lines or adjust line
  depths
\item
  Calculate LSD profiles
\item
  Calculate line bisectors
\item
  Calculate radial velocities
\item
  Calculate \(\langle B_z \rangle\) values
\end{itemize}

The continuum normalization routine follows the algorithm briefly
described in Folsom et al. (\citeproc{ref-Folsom2008}{2008}) and Folsom
(\citeproc{ref-Folsom2013}{2013}), and implements an interactive
graphical interface with the Tkinter and matplotlib packages. Line masks
can be generated from line lists in the
\href{https://vald.astro.uu.se/}{Vienna Atomic Line Database}
(\citeproc{ref-Ryabchikova2015}{Ryabchikova et al., 2015}) ``extract
stellar'' ``long'' format. If necessary, effective Landé factors are
estimated using LS, J\(_1\)J\(_2\), and J\(_1\)K coupling schemes
(\citeproc{ref-LandiAndLandolfi2004}{Landi Degl'Innocenti \& Landolfi,
2004}; \citeproc{ref-Martin1978}{Martin et al., 1978}). The LSD
calculation follows the method of Donati et al.
(\citeproc{ref-Donati1997}{1997}), with details from Kochukhov et al.
(\citeproc{ref-Kochukhov2010}{2010}). It relies on
\href{https://numpy.org/}{numpy} and makes careful use of numpy's sparse
arrays for efficiency. The interactive line mask cleaning tool uses
Tkinter and matplotlib for the interface. This tool can remove lines
from the mask, and automatically fit line depths using a reference
observed spectrum, following Grunhut et al.
(\citeproc{ref-Grunhut2017}{2017}) with some optimizations. The line
depth fitting routine inverts the linear least squares problem in LSD,
and instead fits line depths given an observation and LSD profile. That
LSD profile must be approximately correct, calculated using a mask (or
part of a mask) with dominantly acceptable lines. Line depth fitting
remains a weighted linear least squares problem, similar to the LSD
calculation, and can be solved efficiently using spare matrix
operations. Radial velocities are calculated by fitting a Gaussian to an
LSD profile by default, although calculating from first moments is also
supported. The \(\langle B_z \rangle\) calculation uses the first moment
technique applied to LSD profiles (\citeproc{ref-Donati1997}{Donati et
al., 1997}; \citeproc{ref-Kochukhov2010}{Kochukhov et al., 2010};
\citeproc{ref-Rees1979}{Rees \& Semel, 1979}).

SpecpolFlow enables users to build their own custom workflow from the
available classes and functions, which can enhance scientific archiving
and reproducibility. This is also valuable for training or involving
students in projects. \texttt{SpecpolFlow}'s
\href{https://folsomcp.github.io/specpolFlow/}{website} includes an
extensive set of tutorials that explain the use of the package and
demonstrate some common analysis cases. These tutorials, in the form of
\href{https://jupyter.org/}{Jupyter Notebooks}, can flexibly be used
within a classroom or workshop setting.

\section{Acknowledgements}\label{acknowledgements}

The authors gratefully acknowledge contributions to the early
development and testing of \texttt{SpecpolFlow}'s tools, including Dax
Moraes, David Meleney Jr., and Gregg Wade.

This research was supported by the Munich Institute for Astro-, Particle
and BioPhysics (MIAPbP), which is funded by the Deutsche
Forschungsgemeinschaft (DFG, German Research Foundation) under Germany's
Excellence Strategy -- EXC-2094 -- 390783311.

CPF gratefully acknowledges funding from the European Union's Horizon
Europe research and innovation programme under grant agreement
No.~101079231 (EXOHOST), and from the United Kingdom Research and
Innovation (UKRI) Horizon Europe Guarantee Scheme (grant number
10051045).

VP, SS, PS, and TN gratefully acknowledge support for this work from the
National Science Foundation under Grant No.~AST-2108455.

MEO gratefully acknowledges support for this work from the National
Science Foundation under Grant No.~AST-2107871.

SS gratefully acknowledges support for this work from the Delaware Space
Grant College and Fellowship Program (NASA Grant 80NSSC20M0045).

BZ acknowledges funding from the CAPES-PrInt program
(\#88887.683070/2022-00 and \#88887.802913/2023-00).

The \texttt{SpecpolFlow} team also thanks Ms.~Tali Natan for her
creative design of the SpecpolFlow logo.

\section*{References}\label{references}
\addcontentsline{toc}{section}{References}

\phantomsection\label{refs}

\end{document}